\documentclass{ws-procs9x6}

\begin{document}

\title{Axions and their Distribution in Galactic Halos}

\author{P.~SIKIVIE}

\address{Department of Physics, \\
University of Florida, \\
Gainesville, FL 32611-8440 \\
E-mail: sikivie@phys.ufl.edu}


\maketitle

\abstracts{ Axion physics is briefly reviewed.  Constraints from
laboratory searches, astrophysics and cosmology require the axion 
mass to be in the range $10^{-6} \lesssim m_a < 3\cdot 10^{-3}$eV.  
Near the lower end of this range, axions are all or a major component
of the cold dark matter of the universe.  The late infall of axions, 
and of any other cold dark matter particles, onto our galaxy produces 
streams and caustics in its halo.  The outer caustics are topological 
spheres whereas the inner caustics are rings.  The self-similar model 
of galactic halo formation predicts that the caustic ring radii $a_n$
obey the approximate law $a_n \sim 1/n$.  Evidence for this law has 
been found in a statistical study of 32 extended and well-measured
external galactic rotation curves, and in the existence and distribution 
of sharp rises in the Milky Way rotation curve.  Moreover, a triangular 
feature in the IRAS map of the Galactic plane is consistent with the 
imprint of a ring caustic upon the baryonic matter.  Its position
coincides with a rise in the rotation curve, the one nearest to us.  
These observations imply that the dark matter in our neighborhood is 
dominated by a single flow.  Estimates of that flow's density and 
velocity vector are given.}

\section{Axions}
 
The axion was postulated approximately 25 years ago \cite{arev} to 
explain why the strong interactions conserve P and CP in spite 
of the violation of those symmetries by the weak interactions.  The 
QCD Lagrangian
\begin{eqnarray}
L_{QCD} = -{1\over 4} G^a_{\mu\nu} G^{a\mu\nu} &+& \sum_{j=1}^n
\left[ \overline q_j \gamma^\mu i D_\mu q_j 
- (m_j q_{Lj}^\dagger q_{Rj} + \hbox{h.c.})\right]\nonumber\\
&+& {\theta g^2\over 32\pi^2} G^a_{\mu\nu} \tilde G^{a\mu\nu} ~~\ .
\label{QCD}
\end{eqnarray}
violates P and CP unless $\theta = 0~({\rm mod}~\pi)$.  The absence of 
P and CP violations in the strong interactions, in particular the 
experimental bound on the neutron electric dipole moment, yields the 
limit: $\theta < 10^{-9}$.  However, in the Standard Model of particle 
physics, because of P and CP violation by the weak interactions,
one would expect $\theta$ to be of order one. This discrepancy is 
called the 'strong CP problem'.

The existence of an axion solves this problem in a simple manner which 
is rich in implications for experiment, for astrophysics and for
cosmology.  In axion models, the $\theta$-parameter in Eq. (\ref{QCD})
is replaced by 
\begin{equation}
\theta_{\rm eff} = \theta + {a(x) \over f_a}
\label{repl}
\end{equation}
where $a(x)$ is the axion field and $f_a$, called the axion decay
constant, is of order the energy scale at which the U(1) symmetry of
Peccei and Quinn is spontaneously broken.  The axion is the 
quasi-Nambu-Goldstone boson associated with this symmetry breaking.  
In axion models, the effective value of $\theta$, Eq. (\ref{repl}),
relaxes to zero dynamically. The axion mass is given in terms of 
$f_a$ by
\begin{equation}
m_a\simeq 6~\mu{\rm eV}~{10^{12} {\rm GeV}\over f_a}\, .
\end{equation}
The axion has zero spin, zero electric charge, and negative 
intrinsic parity.  All its couplings are inversely proportional 
to $f_a$.  Of particular relevance to present axion searches is 
its coupling to two photons:
\begin{equation}
L_{a\gamma\gamma} = -g_\gamma {\alpha\over \pi} {a(x)\over f_a}
\vec E \cdot\vec B
\end{equation}
where $\vec E$ and $\vec B$ are the electric and magnetic fields,
$\alpha$ is the fine structure constant, and $g_\gamma$ is a
model-dependent coefficient of order one.  

The axion mass is not known a-priori.  Indeed the axion solves 
the strong CP problem for any value of its mass.  However masses 
larger than 50 keV are ruled out by searches for the axion in
high energy and nuclear physics experiments.  Also, masses between 
300 keV and 3 milli-eV are ruled out by stellar evolution, 
specifically the ages of red giants and the duration of the 
observed neutrino pulse from Supernova 1987a.  Finally, masses 
less than about one micro-eV are ruled out because axions that 
light would be so abundantly produced in the early universe as 
to exceed the closure density.  In summary, the only remaining 
window of allowed axions masses is 
$10^{-6} \lesssim m_a < 3\cdot 10^{-3}$eV.

In that window, axions contribute importantly to the present 
cosmological energy density.  Axions are one of the leading 
cold dark matter candidates.  Dark matter axions can be searched 
for on Earth by stimulating their conversion to micro-wave
photons in an electromagnetic cavity permeated by a strong 
magnetic field \cite{axdet}. Searches of this type are presently 
under way in the US \cite{ADMX} and in Japan \cite{carrack}.  The 
method of axion to photon conversion in a magnetic field can also 
be applied to solar axion detection \cite{axdet,KVB}.  A limit was
published this year by Minowa's group \cite{mino} in Tokyo.  Also 
the CAST experiment at CERN \cite{ziou}, using a decommisioned 
LHC magnet, is getting ready to take data.

The ADMX experiment at Lawrence Livermore Laboratory, which searches 
for dark matter halo axions, is described by D. Kinion at this meeting.  
If it discovers a signal, the detector will be able to measure the 
kinetic energy spectrum of cold dark matter with great precision 
and resolution.  The spectrum of axions and WIMPs is the same, in 
the (excellent) approximation where the primordial velocity dispersion 
of both cold dark matter candidates is neglected, because it is the 
outcome of purely gravitational interactions.  My collaborators and 
I have been motivated to try and predict this spectrum.  It has been 
an exciting adventure which I report on in the rest of my talk.

\section{Flows and Caustics of Dark Matter}

The model of the structure of the halos of isolated galaxies which we have
developed is based on the observation that the dark matter particles must
lie on a 3-dimensional sheet in phase-space, that this sheet cannot break,
and hence that its evolution is constrained by topology.  The thickness of
the sheet is the velocity dispersion.  The primordial velocity dispersion
of the leading cold dark matter candidates is extremely small, of order
$10^{-12}c$ for WIMPs and $3\cdot 10^{-17}c$ (at most) for axions. For a
coarse-grained observer the sheet may have additional velocity dispersion
because it is wrapped up on scales which are small compared to the galaxy
as a whole.  This latter effective velocity dispersion is associated with
the clumpiness of the dark matter before it falls onto the galaxy.  The
effective velocity dispersion of the infalling dark matter must be much
less than the rotation velocity of the galaxy for the model to have
validity, say less than 30 km/s for our galaxy. On the other hand, by
comparing the model with observations, an upper bound of order 50 m/s 
has been obtained, as described below.

Where a galaxy forms, the sheet wraps up in phase-space, turning clockwise
in any two dimensional cut $(x, \dot{x})$ of that space.  $x$ is the
physical space coordinate in an arbitrary direction and $\dot{x}$ its
associated velocity.  The outcome of this process is a discrete set of
flows at any physical point in a galactic halo \cite{ips}.  Two flows
are associated with particles falling through the galaxy for the first
time ($n=1$), two other flows are associated with particles falling
through the galaxy for the second time ($n=2$), and so on.  Scattering
in the gravitational wells of inhomogeneities in the galaxy (e.g.
molecular clouds and globular clusters) are ineffective in thermalizing
the flows with low values of $n$.

Caustics appear wherever the projection of the phase-space sheet onto
physical space has a fold \cite{cr,lens,Tre,sing}.  Generically, caustics
are surfaces in physical space.  On one side of the caustic surface
there are two more flows than on the other.  At the surface, the dark
matter density is very large.  It diverges there in the limit of zero
velocity dispersion.  There are two types of caustics in the halos of
galaxies, inner and outer.  The outer caustics are topological spheres
surrounding the galaxy.  They are located near where a given outflow
reaches its furthest distance from the galactic center before falling
back in.  The inner caustics are rings \cite{cr}.  They are located
near where the particles with the most angular momentum in a given
inflow reach their distance of closest approach to the galactic center
before going back out.  A caustic ring is a closed tube whose
cross-section is a $D_{-4}$ (also called {\it elliptic umbilic})
catastrophe \cite{sing}.  The existence of these caustics and their
topological properties are independent of any assumptions of symmetry.

Primordial peculiar velocities are expected to be the same for baryonic
and dark matter particles because they are caused by gravitational forces.
Later the velocities of baryons and CDM differ because baryons collide
with each other whereas CDM is collisionless. However, because angular
momentum is conserved, the net angular momenta of the dark matter and
baryonic components of a galaxy are aligned.  Since the caustic rings
are located near where the particles with the most angular momentum in
a given infall are at their closest approach to the galactic center,
they lie close to the galactic plane.

\section{A Formula for the Caustic Ring Radii}

A specific proposal has been made for the radii $a_n$ of caustic rings
\cite{cr}:
\begin{eqnarray}
\{a_n: n=1,2, ...\} \simeq (39,~19.5,~13,~10,~8,...){\rm kpc}\nonumber\\
\times \left({j_{\rm max}\over 0.25}\right) \left({0.7\over h}\right)
\left({v_{\rm rot} \over 220 {{\rm km} \over {\rm s}}} \right)
\label{crr}
\end{eqnarray}
where $h$ is the present Hubble constant in units of
$100\,{\rm km/(s~Mpc)}$, $v_{\rm rot}$ is the rotation velocity of the
galaxy and $j_{\rm max}$ is a parameter with a specific value for each
halo.  For large $n$, $a_n \sim 1/n$.  Eq. \ref{crr} is predicted by
the self-similar infall model \cite{ss,sty} of galactic halo formation.
$j_{\rm max}$ is then the maximum of the dimensionless angular momentum
$j$-distribution \cite{sty}.  The self-similar model depends upon a
parameter $\epsilon$ \cite{ss}.  In CDM theories of large scale structure
formation, $\epsilon$ is expected to be in the range 0.2 to 0.35
\cite{sty}.
Eq. \ref{crr} is for $\epsilon = 0.3$.  However, in the range
$0.2 < \epsilon < 0.35$, the ratios $a_n/a_1$ are almost
independent of $\epsilon$.  When $j_{\rm max}$ values are quoted
below, $\epsilon = 0.3$ and $h = 0.7$ will be assumed.

Since the caustic rings lie close to the galactic plane, they cause
bumps in the rotation curve, at the locations of the rings.  In
ref. [15] a set of 32 extended well-measured rotation curves
was analyzed and statistical evidence was found for bumps distributed
according to Eq. \ref{crr}.  That study suggests that the $j_{\rm max}$
distribution is peaked near 0.27.  The rotation curve of NGC3198, one
of the best measured, by itself shows three faint bumps which are
consistent with Eq. \ref{crr} and $j_{\rm max} = 0.28$.

Because angular momentum has the effect of depleting the inner halo,
an effective core radius is produced when angular momentum is included
into the self-similar infall model \cite{sty}.  The average amount of
angular momentum of the Milky Way halo can be estimated by requiring
that approximately half of the rotation velocity squared at our location
is due to dark matter, the other half being due to baryonic matter.
This yields $\bar{j} \sim 0.2$ where $\bar{j}$ is the average of the
$j$-distribution for our halo \cite{sty}.  If the $j$-distribution is
taken to be that of a rigidly rotating sphere, we have
$j_{\rm max} = {4 \over \pi} \bar{j} \sim 0.254$. The value which will
be obtained below (0.263) from a fit to bumps in the Milky Way rotation
curve is consistent with this earlier estimate.

\section{The Milky Way rotation curve}

Galactic rotation curves are obtained from HI and CO surveys of the
Galactic plane.  A list of surveys performed to date is given in
ref. [16].  The CO surveys have far better angular resolution
than the HI surveys because their wavelength is nearly two orders of
magnitude smaller (0.26 cm vs. 21 cm).  The most detailed Galactic
rotation curve appears to be that obtained \cite{clem} from the
Massachusetts-Stony Brook North Galactic Plane CO survey \cite{CO}.
It is reproduced in Fig. 1.  It exhibits a series of ten sharp rises
\begin{figure}[ht]
%\vspace{1.5cm}
%\psfig{file=fig1.eps,height=6in,width=4.5in}
%\epsfxsize=6.5in
%\centerline{\epsfbox{rotcurv.ps}}
%\vspace{1.5cm}
\centerline{\epsfxsize=4.5in\epsfbox{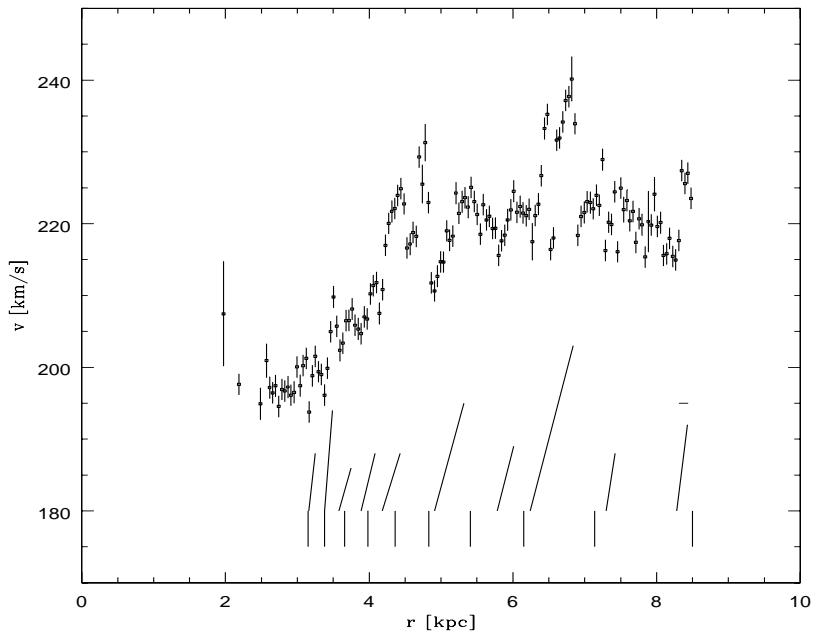}}
\caption{North Galactic rotation curve from ref. [17].  The locations of
the rises are indicated by line segments parallel to the rises but shifted
downwards.  The caustic ring radii in the fit described in the text are
shown as vertical line segments.  The position of the triangular feature 
in the IRAS map of the galactic plane near $80^\circ$ longitude is shown
by the short horizontal line segment.}
\end{figure}
between 3 kpc and our own radius, taken to be 8.5 kpc.

The rises can be interpreted \cite{milk} as due to the presence of
caustic rings of dark matter.  The effect of a caustic ring in the
plane of a galaxy upon its rotation curve was analyzed in ref. [12].
It was shown that the radius $r_1$ where the rise starts should be
identified with the caustic ring radius $a_n$, and the radius $r_2$ 
where the rise ends should be indentified with $a_n + p_n$ where $p_n$ 
is the caustic ring width. The ring widths depend in a complicated way 
on the velocity distribution of the infalling dark matter at last
turnaround \cite{sing} and are not predicted by the model.  They 
also need not be constant along the ring.  The rises between 3 and 
7 kpc were fitted to the model prediction for the caustic ring radii, 
Eq. \ref{crr}.  For $\epsilon = 0.3$ and $j_{\rm max} = 0.263$, one 
has $rmsd \equiv [{1 \over 10}{\displaystyle \sum_{n=5}^{14}
( 1 - {a_n \over r_{1 n}})^2]^{1 \over 2}} = 3\%$.  So we find that, 
for model parameters in the expected range, a sharp rise is present 
in the rotation curve for each of the caustic rings typically within 
3\% of the predicted radius.

In the past, rises (or bumps) in galactic rotation curves have been
interpreted as due to the presence of spiral arms.  However there are
reasons to believe that the rises in the high resolution rotation curve of
Fig. 1 are caused by caustic rings rather than spiral arms. First, there
are of order ten rises in the range of radii covered (3 to 8.5 kpc)
whereas the expected number of spiral arms is much less than ten.  Only
three spiral arms are known in that range: Scutum, Sagittarius and Local.
On the other hand, the observed number of rises agrees with the number of
caustic rings for the expected values \cite{sty} of the model parameters,
e.g. $\epsilon = 0.3$ and $j_{\rm max} = 0.254$.  Second, the rises are
sharp transitions in the rotation curve, both where they start $(r_1)$and
where they end $(r_2)$.  One would expect spiral arms to produce smoother
features.  Sharp transitions are consistent with caustic rings because the
latter have divergent density at $r_1 = a$ and $r_2 = a + p$ in the limit
of vanishing velocity dispersion.  Finally, there are bumps and rises in
rotation curves measured at galactocentric distances much larger than the
disk radius, where there are no spiral arms seen.  In particular, the
features found in the composite rotation curve constructed in ref. [9] are
at distances 20 kpc and 40 kpc when scaled to our own galaxy.  So there
already exists evidence that some bumps or rises are not due to spiral
arms.

\section{Caustic Ring Imprint}

The location of caustic rings may be revealed by the gas that they accrete. 
Looking tangentially to a ring caustic from a vantage point in the plane
of the ring, one may recognize the tricusp \cite{sing} shape of the
$D_{-4}$ catastrophe.  I searched for such features.  The IRAS map of the
galactic disk in the direction of galactic coordinates $(l,b) = (80^\circ,
0^\circ)$ shows a triangular shape which is strikingly reminiscent of the
cross-section of a ring caustic.  The vertices of the triangle are at
$(l,b) = (83.5^\circ, 0.3^\circ), (77.3^\circ, 3.5^\circ)$ and
$(77.4^\circ, -2.7^\circ)$.  Images can be obtained from the
Skyview Virtual Observatory (http://skyview.gsfc.nasa.gov/).  The shape
is correctly oriented with respect to the galactic plane and the galactic
center.  To an extraordinary degree of accuracy it is an equilateral
triangle with axis of symmetry parallel to the galactic plane, as is
expected for a caustic ring whose transverse dimensions are small compared
to its radius.  Moreover its position is consistent with the position of a
rise in the rotation curve, the one between 8.28 and 8.43 kpc ($n=5$ in
the fit).  The caustic ring radius implied by the image is 8.31 kpc, and
its dimensions are $p = 130$ pc and $q = 200$ pc, in the directions
parallel and perpendicular to the galactic plane respectively.  It
therefore predicts a rise which starts at 8.31 kpc and ends at 8.44 kpc,
just where a rise is observed.  Even if a rise were expected in the
neighborhood, say randomly placed within a kpc of the triangular shape,
the probability that its position agrees that closely with the position 
of the triangular shape is only of order $10^{-3}$.

In principle, the feature at $(80^\circ, 0^\circ)$ should be matched
by another in the opposite tangent direction to the nearby ring caustic,
at approximately $(-80^\circ, 0^\circ)$.  Although there is a plausible
feature there, it is much less compelling than the one in the
$(+80^\circ, 0^\circ)$ direction.  There are several reasons why
it may not appear as strongly.  One is that the $(+80^\circ, 0^\circ)$
feature is in the middle of the Local spiral arm, whose stellar
activity enhances the local gas emissivity, whereas the
$(-80^\circ, 0^\circ)$ feature is not so favorably located.  Another is
that the ring caustic in the $(+80^\circ, 0^\circ)$ direction has
unusually small dimensions.  This may make it more visible by increasing
its contrast with the background.  In the $(-80^\circ,0^\circ)$ direction,
the nearby ring caustic may have larger transverse dimensions.

Our proximity to a ring means that the associated flows, i.e. those
flows in which the caustic occurs, contribute very importantly to the
local dark matter density.  Using the results of refs. \cite{cr,sing,sty},
and assuming axial symmetry of the caustic ring between us and the
tangent point (approx. 1 kpc away from us), the densities and velocity
vectors on Earth of the associated flows can be derived:
\begin{eqnarray}
d^+ = 1.7~10^{-24}~{{\rm gr} \over {\rm cm}^3}~&,&~
d^- = 1.5~10^{-25}~{{\rm gr} \over {\rm cm}^3}~,~ \nonumber\\
\vec{v}^\pm = (470~\hat{\phi} &\pm& ~100~\hat{r})~
{{\rm km} \over {\rm s}},
\label{lc}
\end{eqnarray}
where $\hat{r}, \hat{\phi}$ and $\hat{z}$ are the local unit vectors
in galactocentric cylindrical coordinates.  $\hat {\phi}$ is in the 
direction of galactic rotation.  The velocities are given in the
(non-rotating) rest frame of the Galaxy.  Because of an ambiguity, 
it is not presently possible to say whether $d^\pm$ are the densities 
of the flows with velocity $\vec{v}^\pm$ or $\vec{v}^\mp$.

Previous estimates of the local dark matter density, based on isothermal
halo profiles, range from 5 to 7.5~$10^{-25}~{{\rm gr} \over {\rm cm}^3}$.
The present analysis implies that a single flow ($d^+$) has three times
that much local density, i.e. that the total local density is four
times higher than previously thought.  The large size of $d^+$ is due
to our proximity to a cusp of the nearby caustic.  Assuming axial
symmetry, that cusp is only 55 pc away from us.  The exact size of
$d^+$ is sensitive to our distance to the cusp but, in any case,
$d^+$ is very large.  If we are inside the tube of the fifth caustic,
there are two additional flows on Earth, aside from those given in
Eq. \ref{lc}.  A list of approximate local densities and velocity
vectors for the $n \neq 5$ flows can be found in ref. [21]. An updated
list is in preparation.

The sharpness of the rises in the rotation curve and of the triangular
feature in the IRAS map implies an upper limit on the velocity dispersion
$\delta v_{\rm DM}$ of the infalling dark matter.  Caustic ring
singularities are spread over a distance of order
$\delta a \simeq {R~\delta v_{\rm DM} \over v}$ where $v$ is the velocity
of the particles in the caustic, $\delta v_{\rm DM}$ is their velocity
dispersion when they first fell in, and $R$ is the turnaround radius
then.  The sharpness of the IRAS feature implies that its edges are
spread over $\delta a \lesssim 20$ pc.  Assuming that the feature is
due to the $n=5$ ring caustic, $R \simeq$ 180 kpc and $v \simeq 480$
km/s.  Therefore $\delta v_{\rm DM} \lesssim 53$ m/s.

The caustic ring model, and more specifically the prediction Eq.~\ref{lc}
of the locally dominant flow associated with the nearby ring, has
important consequences for axion dark matter searches \cite{adm},
the annual modulation \cite{bux,ann,wick} and signal anisotropy
\cite{anis,stiff} in WIMP searches, the search for $\gamma$-rays
from dark matter annihilation \cite{gam}, and the search for
gravitational lensing by dark matter caustics \cite{lens,our}.  
The model allows precise predictions to be made in each of these
approaches to the dark matter problem.

\section*{Acknowledgments}
This work was supported in part by the US Department of Energy 
under grant No. DE-FG02-97ER41029.

\end{document}